\documentclass[aps,prl,twocolumn,noshowpacs,superscriptaddress,groupedaddress]{revtex4} 
\usepackage{graphicx}  
\usepackage{dcolumn}   
\usepackage{bm}        
\usepackage{amssymb}   

\usepackage[T2A]{fontenc}

\usepackage[utf8]{inputenc}

\usepackage[english,russian]{babel}

\usepackage{amsmath}

\usepackage{amsfonts}

\usepackage{textcomp}

\usepackage{bm}

\usepackage{color}

\usepackage{verbatim}

\usepackage{subfig}

\begin{document}

\title{Chiral vortical effect for vector fields}

\author{G. Yu. Prokhorov}
\email{prokhorov@theor.jinr.ru}
\affiliation{JINR, Dubna, Russia}
\author{O.V. Teryaev}
\email{teryaev@jinr.ru}
\affiliation{JINR, Dubna, Russia}
\affiliation{ITEP, B. Cheremushkinskaya 25, Moscow, 117218 Russia}
\author{V.I. Zakharov}
\email{vzakharov@itep.ru}
\affiliation{ITEP, B. Cheremushkinskaya 25, Moscow, 117218 Russia}
\affiliation{School of Biomedicine, Far Eastern Federal University, Vladivostok 690950, Russia}

\begin{abstract}
We consider photonic vortical effect, i.e. the difference of the flows of left- 
and  right-handed photons along the vector of angular velocity
in rotating photonic medium. Two alternative
frameworks to evaluate the effect are considered, both of which have already
been tried in the literature. First, the standard thermal fied theory
and, alternatively, Hawking-radiation-type derivation. In our earlier attempt
to compare the two approaches, we found a crucial factor of two difference. 
Here we revisit
the problem, paying more attention to details of infrared regularizations.
We find out that introduction of an infinitesimal mass of the vector field
brings the two ways of evaluating the chiral vortical effect 
into agreement with each other. Some implications, both on the theoretical 
and phenomenological sides, are mentioned.
\end{abstract}

\maketitle
\section{Introduction}

We will consider 
thermodynamics of media whose constituents are massless
particles of non-zero spin $S$.
 The best studied case 
is
$S=1/2$ and , as a starting point we quote some 
results obtained for spin 1/2 constituents.
However, we are mainly interested in properties of photonic media, 
consisting of left- and right-handed photons. Since recently, 
the number of papers devoted to this case has also been growing,
see in particular
  \cite{songolkar,ren,agullo,sadofyev1,sadofyev2,landsteiner,yamamoto,
prokhorov1,mitkin}.

The chiral vortical effect was first evaluated by A. Vilenkin 
\cite{vilenkin} 
who considered
gas of non-interacting spin-1/2 fermions in a rotating coordinate system.
It was demonstrated that there exists a current of the particle number 
flowing along the vector of
the  angular velocity $\vec{\Omega}$. Numerically, 
in case of a single right-handed Weyl spinor the current is given by:
\begin{equation}\label{cve}
\vec{J}_N(S=1/2)~=~\Big(\frac{\mu_R^2}{4\pi^2}+\frac{T^2}{12}\Big)\vec{\Omega}~,
\end{equation}
where 
$\mu_R$ is the chemical potential, and $T$ is the temperature.

Nowadays, the literature on the chiral effects is huge and we cannot even briefly 
review the subject. Here we mention only the pioneering paper \cite{sonsurowka}
which opened the chapter on theory of the chiral effects in the regime
of strong coupling. It turns out that in the hydrodynamic approximation and
in absence of dissipation one can derive chiral effects without 
exploiting the non-interacting gas approximation.
Moreover, the magnitude of the chiral effects is determined uniquely
in terms of the corresponding chiral anomaly of the fundamental theory 
underlying
the phenomenological hydrodynamic approach, for review see the volume
\cite{reviews}.

In particular, in case of the chiral vortical effect (\ref{cve}) the term proportional to the
chemical potential squared is indeed related to the chiral anomaly. There is a
simple substitution 
which allows to generate chiral hydrodynamic effects from the standard
chiral anomaly \cite{shevchenko}:
\begin{equation}\label{substitution}
eA_{\alpha}~\to~eA_{\alpha}+\mu u_{\alpha}~~,
\end{equation}
where $A_{\alpha}$ is the electromagnetic potential, $e$ is the charge 
of the fundamental 
constituents and $u_{\alpha}$ is the 4-velocity of an element of the medium.
One can readily check that the substitution (\ref{substitution}) does reproduce
the $\mu_R^2$ term in the Eq. (\ref{cve}).

On the other hand, any field-theoretic interpretation of the $T^2$ term in the Eq. 
(\ref{cve}) had been missing until the recentl paper \cite{stone}
which relates it to the gravitational chiral anomaly (for earlier attempts 
in the same direction see \cite{megias,megias1}). In case of massless spin 1/2
particles interacting with external gravitational field the anomaly reads:
\begin{equation}\label{gravity}
\nabla_{\mu}J_N^{\mu}~=~
-\frac{1}{384\pi^2}R_{\mu\nu\kappa\lambda}\tilde{R}^{\mu\nu\kappa\lambda}~~,
\end{equation}
where 
$R_{\mu\nu\kappa\lambda}$ is the Riemann tensor,
$\tilde{R}^{\mu\nu\kappa\lambda}~=~\frac{1}{2}\epsilon^{\mu\nu\rho\sigma}
R_{\rho\sigma}^{\kappa\lambda}$.

To bridge (\ref{gravity}) to the chiral vortical effect (\ref{cve}) one exploits
the construction similar to the one 
introduced first to relate the Hawking radiation
from a black hole to the field-theoretic anomalies
\cite{wilzcek}. Namely, one considers a space-time with
a horizon.  What is specific for  the horizon is that
there is a flow of particles from the horizon and absorption 
of the particles falling onto the horizon.  One can say,
there is a right-left asymmetry at the horizon. The rate
of the particle production at the horizon 
can be fixed in terms of the anomalies
of the field theory, or of the gravitational field on the horizon. In more detail, 
the relevant anomaly looks as::
\begin{equation}
\nabla_{\mu}T^{\mu\nu}~=-\frac{c}{96\pi}\frac{\epsilon^{\nu\alpha}}{\sqrt{|g|}}
\partial_{\alpha} R~,
\end{equation}
 where $\epsilon^{\mu\alpha}$ is 
the 2d antisymmetric tensor, $c$ is the central charge and $R$ is the Ricci scalar.
On the other
hand, far off from the horizon the flow of the particles can be compared
to the thermal radiation. It was demonstrated \cite{wilzcek} that the matching
of the two expressions for the flow
of the particles  reproduces the Hawking, or Unruh temperature. 
The calculation is to be performed for each spherical wave separately
and cumbersome technically.

The paper in Ref. \cite{stone} evaluates the chiral vortical effect in a similar way.
There is a significant simplification, however. Namely,
the metric introduced is not a solution of the Einstein equations
but rather imitates rotation of a fluid:
\begin{eqnarray}\label{metric}\nonumber
ds^2~=~-f(z)\frac{(dt-\Omega r^2d\phi)^2}{(1-\Omega^2 r^2)}+\frac{1}{f(z)}dz^2\\
+ dr^2~+~\frac{r^2(d\phi-\Omega dt)^2}{(1-\Omega^2 r^2)}.
\end{eqnarray}
At large distances $z$ the function $f(z)$ tends to unit, and the metric (\ref{metric})
reduces to that of the flat space in cylindrical coordinates:
\begin{equation}
ds^2~\to~-dt^2+dz^2+dr^2+r^2d\phi^2~~.
\end{equation}
 Integrating the r.h.s. of the Eq. (\ref{gravity}) one evaluates
the flow of particles at large $z$ which is to be identified with
the chiral thermal vortical effect, by the logic of the construction.
The translation from the field theory to the thermal physics
is achieved through
the identification:
\begin{equation}
\frac{a_{horizon}}{2\pi}~\to ~T~,~~~~\Omega_{horizon}~\to~\Omega~,
\end{equation}
where $a_{horizon}$ and $ \Omega_{horizon}$ are the gravitational acceleration
and angular velocity
on the horizon, while $T$ and $\Omega$ are the flat-space values of the temperature
and of angular velocity, respectively, as measured at large $z$. see eq. (\ref{metric}).

It was demonstrated \cite{stone} that 
in this way one reproduces the $T^2$ term in the
Eq. (\ref{cve}). Which is an amusing success of the modern ideas on
the relation between gravitational acceleration and thermal physics.
 
In Ref. \cite{prokhorov1} it was suggested to extend the 
checks of the theory \cite{stone}
by considering quantum particles of higher spin $S$. 
In particular, we concentrate on the $S=1$
case, or photons. The corresponding gravitational anomaly
was introduced in Ref. \cite{vainshtein}. Using the machinery just described
one can turn the knowledge of the gravitational anomaly into a 
prediction of the magntude of the chiral vortical effect for photons.
Moreover this predition can be compared with the results of 
direct calculations by means of various techniques within 
the thermal field theory, see in particular 
\cite{ren,songolkar,sadofyev1,agullo,sadofyev2,
landsteiner, yamamoto}. 

There is a disagreement of a factor
of two between the two ways of evaluating the chiral vortical effect for photons 
\cite{prokhorov1}. 
Here we revisit the problem of comparing various results
for the photonic vortical effect. The main point we are 
emphasizing now  is that both the evaluation of the gravitational  anomaly 
and of the chiral
vortical effect involve regulaization procedures. For our, pure theoretical
 purposes,
we need identical regularizations on the both sides  (gravitational and
flat-space ones). 
Our overall conclusion here is that, in the sense indicated,
 there is no direct contradiction between the two
ways of evaluating the chiral vortical effect for photons.

\section{Chiral vortical effect in the equilibrium}

\subsection{Massless spin-1/2 particles}

There are different ways of evaluating  the chiral vortical effect
in the one-loop approximation. The most straightforward way is 
to find energy levels, evaluate the current for each mode and weight
the results with the Fermi or Bose (whichever is relevant) distribution
of the levels. In particular, this was the strategy adopted first in the pioneering work
\cite{vilenkin}. In this section we review briefly evaluation of the photonic chiral 
effect in the equilibrium. In the next section we comment on the applications
of the Kubo relations.

We begin with quoting the results obtained first in \cite{vilenkin} in the form 
most suitable for generalizaions to higher-spin cases. 
For the sake of normalization,
let us remind the reader that the Fermi distribution for a single 
Weyl fermion is written as:
\begin{equation}\label{three1}
n_F^{Weyl}~=~
\frac{1}{8\pi^3}\int_{-\infty}^{+\infty} d\epsilon(4\pi\epsilon^2)
\Big(\frac{1}{e^{\beta\epsilon}+1}-\theta(-\epsilon)\Big)~~.
\end{equation} 
The degree of freedom with a positive energy,
describes a fermion polarized along its momentum.
The second, subtraction term in the r.h.s. of Eq. (\ref{three1})
is introduced to ensure vanishing of the density of the energy at $T,\mu_R=0$.
Alternatively, the Fermi distribution (\ref{three1}) can be written in
terms of states with positive energies, which unifies particles and anti-particles:
\begin{equation}\label{twodof}
n_F^{Weyl}~=~2\frac{1}{8\pi^3}\int_0^{\infty}d\epsilon 
(4\pi\epsilon^2)
\frac{1}{e^{+\beta\epsilon}+1}
\end{equation}
where we keep the overall factor of 2 in the r.h.s. to emphasize that
there are two degenerate levels for each energy.

Now, the statistically averaged matrix element
of the current can be represented
in the following form \cite{vilenkin,stone} convenient for interpretation:
\begin{eqnarray}\label{eighteen}
J_N(s=1/2)~=~\frac{1}{4\pi^2}\int_{-\infty}^{+\infty}\epsilon^2d\epsilon
\Big(\frac{1}{1+e^{\beta(\epsilon-(\mu_R+\Omega/2))}} ~-~\\\nonumber
\frac{1}{1+e^{\beta(\epsilon-(\mu_R-\Omega/2))}}\Big)~,
\end{eqnarray}
where $J_N$ is written, as usual 
for the case of a single right-handed Weyl fermion,
and we restore a non-vanishing chemical potential, $\mu_R\neq 0$.
Moreover,
the first term in the parentheses represents contribution of particles
while the second term refers to the anti-particles. 
Upon integration over the energy, 
Eq. (\ref{eighteen}) reduces to the Eq. (\ref{cve}).

The interpretation of the Eq. (\ref{eighteen}) is straightforward.
Indeed, by introducing the rotation we remove the two-fold degeneracy
of all the levels and get two levels split by the energy $\Delta E=\Omega$.
Indeed, in the equilibrium one introduces an effective interaction:
\begin{equation}\label{effective}
\delta \hat{H}_{eff}~=~\vec{\Omega}\cdot \hat{\vec{M}}~,
\end{equation}
where $\hat{\vec{M}}$ is the angular momentum operator for spin S=1/2.
Note that such an interpretation assumes that the quantization is performed
in the cylindrical coordinates (while the Eq. (\ref{three1}) can be derived, say,
in the Cartesian coordinates as well). Therefore the energy 
levels now correspond to the
states which have a definite projection of the momentum on z-axis, 
$p_z$ and projection
of the angular momentum, $L_z$. 

\subsection{Chiral photonic current}

The notion of chirality for photons is well known. Namely,
 the left- and right-handed
polarized photons are chiral states since they correspond to a 
certain projection of
the spin of the photon on its momentum, $S_p=\pm 1$.
The chiral current, therefore,  can be defined as 
\begin{equation}\label{k1}
K^{\mu}~=~-\frac{1}{\sqrt{-g}}\epsilon^{\mu\nu\rho\sigma}A_{\nu}\partial_{\rho}A_{\sigma}
\end{equation}
where $A_{\mu}$ is the vector potential of the electromagnetic field,
and we also reserved for a non-trivial determinant, $g$ of the metric tensor.
The current (\ref{k1}) is defined in such a way that the eigenvalues 
of the associated charge
are indeed $\pm 1$.

Note, however, that the current (\ref{k1}) is not gauge invariant,
and we should be careful to associate observables only with a kind
of gauge-invariant projections of $K^{\mu}$.
In particular, a well-known example of such a gauge-invariant
observable is the charge coresponding to the current (\ref{k1}):
\begin{equation}
Q_{magnetic~helicity}~=~\int d^3x \epsilon^{0ijk}A_i\partial_jA_k~=~\int d^3x
\vec{H}\cdot\vec{A}~,
\end{equation}
where $\vec{H}$ is the magnetic field. Then,
under the gauge transformation, $\delta_{gauge} A_i~=~ \partial_i\Lambda$, 
the variation of the 
magnetic-helicity charge density is given by:
\begin{equation}
\delta_{gauge} \big(\vec{H}\cdot \vec{A}\big)~=~\vec{H}\cdot\vec{\nabla}\Lambda~=~
\vec{\nabla}\big(\vec{H}\Lambda\big)~-~(\vec{\nabla}\cdot \vec{H})\Lambda.
\end{equation}
Here $\vec{\nabla}\cdot {\vec{H}} = 0$ by virtue of the equations of motion while the integral
over $d^3x$ from the first term becomes a boundary term and can be neglected.
This completes the proof that the $Q_{magnetic~helicity}$ is gauge invariant.
Note that in the momentum space definiton of the charge assumes $q_i\equiv 0,
q_0\to 0$ limiting procedure (where $(q_0,q_i)$ is the 
4-momentum carried by the current).

Note also that the matrix element of the operator of the magnetic-helicity
charge, $\hat{Q}_{magnetic~helicity}$ counts the difference between the numbers
of the left- and right-handed photons 
$n_{L,R}$:
\begin{equation}
<|\int d^3x \epsilon^{0ijk}A_i\partial_jA_k |>~=~n_R-n_L~~.
\end{equation}
In other words, the normalization of the chiral currents for spins $S=1/2,1$
is similar.

Now, to evaluate the photonic vortical effect we need to consider 
the spatial component, $\vec{K}$ of the current (\ref{k1}). Moreover,
since we are interested in the physics of equilibrium we have to consider
the static (or stationary) limit with no time dependence. In the momentum
space, as is first emphasized by \cite{megias1}, we are interested
in the limiting procedure, $q_0\equiv 0, q_i\to 0$. Let us check
that the current $\vec{K}~=~A_0\vec{H}+\vec{A}\times\vec{E}$ 
is gauge invariant in this limit. Under the gauge transformation:
\begin{equation}
\delta_{gauge}\vec{K}~=~\big(\partial_0\Lambda\big)\vec{H}+
\big(\vec{\nabla}\Lambda \big)\times{\vec{E}}~~.
\end{equation}
As a result, we get the local term proportional to
$$\delta_{gauge}\vec{K}~\sim~\Lambda\cdot\big(\partial_0{\vec{H}}-
\vec{\nabla}\times{\vec{E}}\big)$$
which vanishes because of the equations of motion, plus total derivatives which 
become boundary terms upon the integration over the
volume, or time.

To summarize, the apparent gauge-dependence of the chiral photonic current
does not imply, generally speaking, that the chiral photonic current in the equilibrium
is gauge dependent. There is a reservation, however, that the problem considered
is infrared sensitive. In particular, considering uniform rotation everywhere in the space
is inconsistent with finiteness of the speed of light, for discussion see, for example,
\cite{landsteiner}. Thus, neglecting the boundary terms just discussed above
might be in conflict with some other constraints on the behaviour of the fields on the 
boundaries.     

It might worth emphasizing that the current (\ref{k1}) is not a Noether current
and its conservation is not automatic: 
\begin{equation}\label{two2}
\nabla_{\mu}K^{\mu}~=~-\frac{1}{2}F_{\alpha\beta}\tilde{F}^{\alpha\beta}
\end{equation}
where
$\tilde{F}^{\mu\nu}~=~
\frac{1}{2\sqrt{-g}}\epsilon^{\mu\nu\alpha\beta}F_{\alpha\beta}$
However, on the mass shell, or for the electromagnetic
plane waves the r.h.s. of Eq. (\ref{two2}) vanishes, and the 
current (\ref{k1}) is conserved for free photons. 
What is even more fascinating, one can introduce non-trivial dynamics
through interaction
of the photons with external gravitational fields, and the chiral photonic
current is still (naively) conserved. We come back to discuss this point further 
later.

\subsection{Photonic vortical effect}

After these preliminary remarks we are set to consider a direct, {\it a la Vilenkin}
evaluation of the chiral vortical effect for photons. There are no general reasons to
believe that such a calculation  is less reliable than the showcase
\cite{vilenkin,stone}
of spin-1/2 massless fermions. Probably, 
the naive expectation would be
that the final result for the photons is very similar to
(\ref{cve}), with some obvious changes (that is, increasing the splitting,
due to the rotation,  between the levels by a factor of two, due to the spin of the
photon, and replacing the Fermi distribution by the Bose distribution).
The hard work of quantizing photons in the cylindrical coordinates, finding
the levels and the corresponding wave functions, evaluating the magnetic-helicity
current on the modes has been done in Ref. \cite{landsteiner},
with the following  result:
\begin{eqnarray}\nonumber\label{spherical4}
|\vec{K}|~=~{\bf 2/3}\frac{1}{8\pi^3}\int_0^{\infty}d\epsilon 
(4\pi\epsilon^2)\\\Big(\frac{1}{e^{+\beta(\epsilon-\Omega)}-1}
-\frac{1}{e^{+\beta(\epsilon+\Omega)}-1}\Big)~~.
\end{eqnarray}
Almost everything looks like what we expected to find naively. 
Except for the bald-faced
overall factor of 2/3. Thus, in the
rest of this section we will look for a convincing interpretation of this factor.
For further discussion of the relation between the cases of massless and nearly
massless photons see also Appendix

We introduced the current $K^{\mu}$, with an idea that it provides us with 
a unique 
definition of the chirality of the photon. However, it is actually well known 
that there exist various currents which in case of massless particles measure
their spin or chirality,
see  \cite{vainshtein}, and references therein.

The best known spin-related 4-vector has been introduced by Pauli and Lubanski
\cite{lubanski}. This vector, $\Gamma^{\mu}$
 is defined in the most general terms of the 
generators of the Poincare group: 
\begin{equation}
\Gamma^{\mu}~=~\epsilon^{\mu\nu\kappa\lambda}P_{\nu}M_{\kappa\lambda}
\end{equation}
where  $P_{\nu}$ и $M_{\kappa\lambda}$ are the generators
of the coordinate shifts and of the Lorentz rotations,
respectivrly. For the one-partilce state of a massless field, $p^2=0$, of spin $S$
we have:
\begin{equation}
\Gamma_{\mu}|p_{\alpha},\lambda>~=~p_{\mu}\lambda|p_{\alpha},\lambda>~,
\end{equation}
where $\lambda$ is the helicity $\lambda=\pm S$.
The helicity plays a crucial role because it
determines the energy splitting induced by the rotation.
Indeed, we already quoted the expression (\ref{effective}) for
the effective interaction introduced in the statistical physics to describe
the equilibrium.
 Once we quantize the projection of the angular momentum onto
the vector $\vec{\Omega}$ the energy induced by the
effective interaction becomes equal to:
\begin{equation}
\delta E~=~\lambda \Omega~~,
\end{equation}
where $\lambda$ is the helicity.

In view of this, it is useful to introduce a current  $j^{\mu}_{helicity}$,
such that the corresponding charge is equal to the helicity
of the one-particle states.
Explicit expression for the $j^{\mu}_{helicity}$ is:
\begin{equation}
j^{\mu}_{helicity}~=~-\frac{i}{3!}\epsilon^{\mu\nu\rho\sigma}S_{\nu\rho\lambda}~~,
\end{equation}
where $S_{\nu\rho\lambda}$ is the spin part of the density of the angular 
momenum density in the Lagrangian formaism:
\begin{equation}
S_{\nu\rho\lambda}~=~
\frac{\delta L}{\delta (\partial^{\nu}\phi^a)} (\Sigma_{\rho,\lambda})_{ab}\phi^b
\end{equation}
where
$(\Sigma_{\rho\lambda})$ is the represntation of the generators of the Lorentz rotations
on the fields
$\phi_a$:
\begin{equation}
(\Sigma_{\rho\lambda})_{ab}~=~i(g_{\rho a}g_{\lambda b}-g_{\rho b}g_{\lambda a}).
\end{equation}
The crucial point is that for one-particle massless states 
the values of $K^{\mu}$ and of $j^{\mu}_{helicity}$
are proportinal to each other but are not identical.
In particular for a 4-vector field:
\begin{equation}\label{twothird}
(j_{\mu})_{helicity}~=~\frac{2}{3}K_{\mu}~~~~~~4-vector~field~, 
\end{equation}
while in case of the Dirac field the matrix elements of the currents
$j^{\mu}_{helicity}$ and of $K^{\mu}$ coincide with each other.

Apparently, the factors of 2/3 in Eqs (\ref{eighteen}) and (\ref{twothird})
are of pure geometric origin and related to each other. But at the moment
we are not aware of any clear derivation of such a relation. Note also that
these states do not coincide with the propagating states.
To construct propagator in terms of the equilibrium states one needs
to derive re-expansion of one complete set of functions over the other
set. A similar recent analysis can be found in Ref. \cite{becattini2}.

Coming back to our main problem of evaluating the photonic vortical effect,
our next step is the introduction of the infrared regularization 
by a finite photon
mass $m_{\gamma}\neq 0$.

\section{Finite photon mass}
\subsection{Chiral anomaly and infrared regularization}

Introduction of $m_{\gamma}\neq 0$ is a logical step, within our approach.
Indeed, we are going to compare predictions for the photonic vortical effect
obtained in two different ways, namely, in terms of the gravitational chiral anomaly 
for the $K_{\mu}$ current and and  in terms of the statistically averaged matrix
element of the same current. As we remind the reader next, a finite photon mass
is introduced to regularize in the infrared the gravitational anomaly,
\cite{vainshtein}. Therefore, we are invited to consider the statistical-theory approach
at a finite photon mass as well.

To substantiate the point, 
let us reiterate basic steps of derivation of the photonic gravitational anomaly
\cite{vainshtein}.  As we already mentioned, see Eq. (\ref{two2}), there is 
no conservation of the $K_{\mu}$ current off-mass shell. However, for 
electromagnetic waves, or on-mass shell $\vec{E}\cdot \vec{H}~=~0$ and
the current is conserved. It is only natural then that, upon inclusion of interaction
of photons with external gravitational field, we expect covariant conservation,
$\nabla_{\mu}K^{\mu}~=~0$. 

However this, ``naive'' expectation is to be checked against possibility of 
existence of an anomaly. To uncover the anomaly, one considers
the matrix element of transition of the $K_{\mu}$ current into two gravitons,
in the annihilation channel. If the gravitons are on the mass shell the matrix element
is defined in terms of a single form factor $f(q^2)$:
\begin{equation}
<0|K_{\mu}|2g>~=~f(q^2)q_{\mu}R_{\alpha\beta\gamma\delta}
\tilde{R}^{\alpha\beta\gamma\delta}~,
\end{equation}
where  $q_{\mu}$ is the  4-momentum, carried in by the $K_{\mu}$ current.

The next step is to use dispersion relations to evaluate $f(q^2)$. 
The imaginary part, $im f(q^2)$ is given by tree graphs
and, naively, it respects all the symmetries of the problem.
However, a direct calculation of the imaginary part fails
because one of the propagators of intermediate particles
has a pole which--for all the particles being massless--falls onto the
physical region of integration. To regularize the calculation 
one introduces then an infinitesimal photon mass.
As a result:
\begin{eqnarray}\label{imaginary}
Im f(q^2)~=~\lim_{m^2\to 0}\Big(\frac{1}{128\pi q^2}\cdot\\\nonumber
\cdot v^2(1-v^2)\ln\frac{1+v}{1-v}{}\Big) ~=~\frac{1}{96\pi}\delta({q^2})  ~.              
\end{eqnarray}
where 
$v$ is the velocity of the intermediate photons in the c.o.m. system.

Finally,  the real part of $f(q^2)$ corresponding to (\ref{imaginary})
is given by:
\begin{equation}\label{anomaly10}
<\nabla_{\alpha}K^{\alpha}>~=~-\frac{1}{96\pi^2}
R_{\mu\nu\kappa\delta}\tilde{R}^{\mu\nu\kappa\delta}
\end{equation}
where $\tilde{R}^{\mu\nu\kappa\delta}=(1/2)\epsilon^{\kappa\delta\rho\sigma}
R^{\mu\nu}_{\rho\sigma}.$
Eq. (\ref{anomaly10}) is nothing else but the photonic gravitational anomaly.

Following then Ref. \cite{stone} we conclude that the photonic vortical current
is predicted to be:
\begin{equation}\label{ratio}
\vec{J}_N(s=1)~=~4\cdot\frac{T^2}{12}\vec{\Omega}~,
\end{equation}
or four times larger than that for massless spin-1/2 particles
\cite{prokhorov1}.

Turn now to the statistical-theory approach. Introdcution of a
finite photon mass simplifies the evaluation of the chiral vortical effect greatly.
The reason is that, for massive photons, we recover the factorization
property which makes Eq. (\ref{eighteen}) to look so simple.   
Namely, we start with non-interacting gas of massive photons
in absence of the rotation. Each level is degenerate three times
since projection of the spin of the massive photon is now $S_z=\pm1,0$.
Account for the rotation splits the levels so that the energies now
are $\epsilon+\Omega,\epsilon, \epsilon-\Omega$. These energy differences
are readily calculable in the rest frame of the massive photon
and are invariant under the boosts along the
$z$-axis.
In case of massless photons there is no rest frame for photons 
and this makes the calculation much more involved, for further
comments see, in particular, \cite{mitkin}.

Thus, for massive photons the vortical current is gven by:
\begin{eqnarray}\label{statistical}\nonumber
|\vec{K}|_{m_{\gamma}\neq 0}~=~\frac{1}{8\pi^3}\int_0^{\infty} 
d\epsilon (4\pi\epsilon^2)\cdot\\
\cdot\Big(\frac{1}{e^{\beta(\epsilon-\Omega)}-1}-
\frac{1}{e^{\beta(\epsilon+\Omega)}-1}\Big)~=~4\frac{T^2}{12}\Omega,
\end{eqnarray} 
in agreement with the prediction (\ref{ratio}).

This coincidence of the results obtained in case
of massive photons within the thermal field theory and via
the gravitational anomaly is our main result
in these notes. In view of this, we will check it
against calculations of the vortical effect by means of the Kubo relations.

\subsection{Kubo-type relation in case of massive photons}

We are interested to evaluate the coefficient $\sigma_V$ entering
the definition of the vortical current $J^{5\mu}$:
\begin {equation}\label{sigmavort}
J^{5\mu}~=~\frac{\sigma_V}{2}
\epsilon^{\mu\nu\rho\sigma}u_{\nu}\partial_{\rho}u_{\sigma}~,
\end{equation}
where $u_{\mu}$ is the 4-velocity of an element of the fluid.
The Kubo-type relation fixes the coefficient $\sigma_V$ in terms
of the correlator between the spatial components of the current
$K^i$ and the $T^{0j}$ component of the energy-momentum tensor
\cite{megias,songolkar,ren}:
\begin {equation}\label{correlator}
\lim_{p_k\to 0}{<K^i,~T^{0j}>|_{\omega\equiv 0}}=
\sigma_V(S=1)\frac{i}{2}\epsilon^{ijk}p_k+O(p^2)~,
\end{equation}
where $p_k$ is the momentum brought in by the current $K^i$, $K^i=
\epsilon^{i\nu\rho\sigma}A_{\nu}\partial_{\rho}A_{\sigma}$
and
\begin{equation}\label{t0j}
T^{0j}~=~\big(\partial_kA^0-\partial^0A_k\big)\big(\partial^kA^j-\partial^jA^k\big)+
m_{\gamma}^2A^0A^j~~,
\end{equation}
where $A_{\mu}$ is now the field describing massive photons.

The propagator of the massive vector field in the momentum space is given by:
\begin{equation}\label{propagator}
<A_{\mu},~A_{\nu}>~=~\frac{g_{\mu\nu}-\frac{q_{\mu}q_{\nu}}{m_{\gamma}^2}}
{q^2-m_{\gamma}^2}
\end{equation}
In case of massless photons, the correlator (\ref{correlator}) was calculated
in Refs. \cite{songolkar,ren}, with the result quoted above, $\sigma_V(S=1)~=~T^2/6$.
We are calculating now the change in $\sigma_V(S=1)$ due to $m_{\gamma}\neq 0$. It
turns out that in the limit $m_{\gamma}\ll T$ there is a finite jump in the value
of $\sigma_V(S=1)$ which stems from the cancellation of  the factor $m_{\gamma}^{-2}$
in the propagators (\ref{propagator}) and of the factor $m_{\gamma}^2$ in
the  component $T^{0j}$, see Eq. (\ref{t0j}). 

The final result is:
\begin{equation}
\sigma_V(S=1,m_{\gamma}\neq 0)~=~\frac{T^2}{3}~~.
\end{equation}
This result is in full agreement with the expectations (\ref{ratio}) based
on the gravitational anomaly and with the evaluation (\ref{statistical}) of the vortical effect
within the statistical approach for massive photons.   

 \subsection{Vector meson chirality and baryon polarization}

One may ask whether the chirality of massive vector particles has any phenomenological implications. The answer is provided by the relation of axial charge to the average polarization of baryons which is the way to implement the quark-hadron duality in this problem \cite{Sorin:2016smp} which may be realized in both kinetic 
\cite{Baznat:2013zx} and hydrodynamic \cite{Ivanov:2020qqe} calculations. 
There is a natural explanation of the fact the 
$\bar \Lambda $ polarization is larger than that of $\Lambda $ as the same (C-even) chiral charge is distributed between the smaller number of particles. For quantitative description of the effect It is mandatory \cite{Baznat:2017jfj} to take into account the axial charge carried by $K^*$ mesons. Therefore, their chirality is implicitly present here.
The role of the numerical factors studied here depends on the assumptions on the distribution of chirality or axial charge between baryons and mesons and remains to be studied.

It is interesting whether meson chirality can affect the measured tensor polarization of vector mesons \cite{Singha:2020qns}. One should stress that contrary to baryon polarization it is P-even quantity and may emerge due to the product of quark polarizations as well as due to their spin correlations \cite{Efremov:1981vs}:
\begin{equation}
\rho_{00}=\frac{1-Tr_{||} (C) + Tr_{\perp} (C)}{1+3~Tr(C) } 
\end{equation} 
where enter the parallel and orthogonal to quantization axis components of tensor $C$ 
\begin{eqnarray} 
C_{ij}=\langle P^q_i P^{\bar q}_j \rangle = \langle P^q_i \rangle \langle P^{\bar q}_j 
\rangle 
+ {\bf \langle P^q_i P^{\bar q}_j \rangle - \langle P^q_i \rangle \langle P^{\bar q}_j\rangle}
\end{eqnarray}
containng contributions of average quark polarizations and (boldfaced) correlations. The relative smallness of first term is implied by its relation to squared baryon polarization and squared vorticity \cite{Becattini:2020ngo} so that the terms probing the entanglement of quark spins may play the dominant role. 

 At the same time, the role of squared vorticity may be overtaken by square (and higher even powers) of magnetic field \cite{Luschevskaya:2018chr,Luschevskaya:2020aag}. The emerging longitudinal polarization is related \cite{Buividovich:2012kq} to the conductivity in magnetic field 
\cite{Buividovich2010,Astrakhantsev:2019zkr} and supports its growth. 

Let us finally note that vector (related to chirality) and tensor polarizations are mixed in the positivity constraints and invariants of density matrix 
\cite{Artru:2008cp,Gavrilova:2019jea} providing another possible direction of experimental investigations.

\section{Discussion and conclusions}

As is noticed in Ref. \cite{prokhorov1}, knowing the gravitational chiral anomaly 
and following the logic of Ref. \cite{stone} one 
can predict the value of the chiral vortical effect for any spin $S$ of the massless
constituents:
\begin{equation}\label{anyspin}
\sigma_V(S)~=~\frac{T^2}{12}(-1)^{2S}4(2S^3-S) ~,
\end{equation}
where $\sigma_V$ is defined in Eq. (\ref{sigmavort}).
Note that there are no free parameters in this prediction.

Alternatively, one can calculate $\sigma_V$ within the framework
of the thermal field theory and, in this sense, test the Eq. (\ref{anyspin}).
First, and with great success, the prediction (\ref{anyspin}) was tested in 
the original paper \cite{stone} in case of massless spin-1/2 constituents.
This case is remarkable for the fact that  $\sigma_V(S=1/2)$ 
was evaluated in a few ways within the statistical approach,
and the value $\sigma_V(S=1/2)=T^2/(12)$ is well established 
and non-controversial. 

Proceeding to the case $S=1$
\cite{songolkar,ren,agullo,sadofyev1,sadofyev2,landsteiner,yamamoto,
prokhorov1,mitkin} we notice $\sigma_V(S=1)$ 
obtained in the literature \cite{songolkar,ren} on the basis of
the Kubo relation differs from  (\ref{anyspin}) 
by a factor of 2.
Moreover, these results themselves are not without controversy,
see \cite{songolkar,ren,agullo,sadofyev1,sadofyev2,landsteiner,yamamoto,
prokhorov1,mitkin} which is not easy to resolve. Finally, consideration 
of the limit of large spin $S$ apparently brings Eq (\ref{anyspin})
to a qualitative disagreement with the thermal field theory.

It is on this background that we have to appreciate significance of the new 
observation that for a massive vector field there is full agreement of
results for $\sigma_V(S=1,m_{\gamma}\neq 0)$ obtained within the 
field-theoretic and statistical 
approaches. 

As is mentioned above, the main argument against the duality between 
the statistical and anomaly-based approaches is an apparent conflict
between the predictions for the chiral vortical effect obtained within the two 
frameworks in the limit of large spin $S$.
The finding that in case of the vector field 
introduction of   a finite mass $m_{\gamma}\neq 0$  
brings the consequences from the Kubo-type relation and from the anomaly
into agreement with each other does not settle by itself the issue of the
violation of the duality
for large spin $S$.
However,  increasing spin $S$ 
generically  makes the theory
more and more dependent on details of the infrared regularization \cite{prokhorov1}.
In particular, the rotational vacuum becomes unstable at spin $S\ge 3/2$.
This infrared instability does not affect directly the derivation \cite{duff} of the 
graitational anomaly. However, beginning with $S=3/2$ one has to assume that the infrared
issues are settled somehow without changing prediction (\ref{anyspin}) which,
in the language of the thermal field theory, is expected to be saturated
by  contribution of high energies of order temperature.

The results for the chiral vortical effect obtained at $m_{\gamma}
\neq 0$ make the validity of this extra assumption more questionable. 
Indeed, we have demonstrated that introduction of $m_{\gamma}\ll T$
results in a finite jump in the vaue of $\sigma_V(S=1)$. Explicit evaluation 
of $\sigma_V$ in case of $S=3/2$ becomes a crucial step to be made.
(For a recent discussion of the theory of massless charged spin-3/2 particles see
\cite{adler}).  

We conclude this section with a remark on possible phenomenological
implications of the evaluation of $\sigma_V(S=1)$. The point is
that in case of superfluidity the chiral vortical efect can be manifested 
through polarization, or spin of heavy particles (for details and references see
\cite{sonzhitnitsky,zhitnitsky}). In particular, in case of superfluidity
the average value of the vortical current $<\vec{J}^5>$ 
is equal to the spin density carried by
the cores of the vortices, which, in turn, is equal to the spin density
carried by heavy
particles $<\vec{\sigma}_{heavy}>$ \cite{kirilin,teryaevzakharov}:
 \begin{equation}\label{relation}
<\vec{J}^{5}>~\approx ~<\vec{\sigma}_{heavy}>  .
 \end{equation}
In view of the low viscosity of the quark-gluon plasma such a relation
might work well in case of heavy-ion collisions \cite{teryaevzakharov}.
Usually the relation (\ref{relation}) is used in case of spin-1/2 constituents
and applied to hyperons, as heavy particles. Since $\sigma_V(S=1)=4\sigma_V(S=1/2)$
one can speculate that the contribution of heavy mesons is not less important than
the contribution of hyperons.

\section{Acknowlegments}

We are grateful to A. I. Vainshtein and P. G. Mitkin for discussions and critical remarks.
The work was partially supported by the grant RFBR18-02-40056.

\section{Appendix}

In the main body of the paper we 
considered the cases of strictly massless photons and of
photons with infinitesimally small mass. The predictions for the photonic
vortical effect differ by a finite factor of 2/3,
see Eq. (\ref{spherical4}). Nevertheless, it is 
rather obvious that the two results are absolutely consistent with each other.
To appreciate this, we should be more careful to formulate the question
to be answered.

Let us start with no rotation. Then, depending on whether $m_{\gamma}$ is strictly
zero or small we have energy densities which differ from each other by a factor
of 2/3:
\begin{eqnarray}\nonumber
n_{\gamma}(m_{\gamma}\equiv 0)~=~2\frac{1}{8\pi^3}\int_0^{\infty}d\epsilon
(4\pi\epsilon^2)\frac{1}{\exp^{\beta\epsilon}-1}~~\\
n_{\gamma}(m_{\gamma}\neq 0)~=~3\frac{1}{8\pi^3}\int_0^{\infty}d\epsilon
(4\pi\epsilon^2)\frac{1}{\exp^{\beta\epsilon}-1}~~.
\end{eqnarray}
If we switch on the effect of the rotation, the number of levels is not changed.
Moreover the distribution between the levels with $L_{z}=+1,~0~,-1$
apparently is the same for massless photons and photons with infinitesimal
mass (i.e. $m_{\gamma}\ll T)$. For this reason the factor of 2/3 which reflects the 
difference in the total number of levels goes through to the final answer for
the chiral vortical effect. Thus, the factor of 2/3 in Eq. (\ref{spherical4})  reflects
so to say renomalization of the total amount of thermal energy stored in the system.

On the technical side, we argued that the simple, ``factorized'' form of distribution
of levels is valid for massive particles  and is not valid for strictly massless partciles.
This is in accord with the theoretical expectations. The difference vetween massive and 
strictly massless cases goes back to the fact that for massive particles the 4-momentum
vector is ortogonal to the Pauli-Lubanski vector while for strictly massless cases
the two vectors are parallel to each other, for a related discussion see \cite{mitkin}.


\begin{thebibliography}{99}
\bibitem{songolkar}
 S. Golkar and D. T. Son,	
{\it ``(Non)-renormalization of the chiral vortical effect coefficient''}
  JHEP 1502 (2015) 169,
 arXiv:1207.5806 [hep-th] .

\bibitem{ren}
 De-Fu Hou, Hui Liu, and Hai-cang Ren,
{\it   	
``A Possible Higher Order Correction to the Vortical Conductivity 
in a Gauge Field Plasma''}
 Phys. Rev. D86 (2012) 121703, 
arXiv:1210.0969 [hep-th].

\bibitem{agullo}
 I. Agullo, A. del Rio, and J. Navarro-Salas,	
{\it ``Electromagnetic duality anomaly in curved spacetimes''},
  Phys. Rev. Lett. 118 (2017), 111301,
  arXiv:1607.08879 [gr-qc].

\bibitem{sadofyev1} 
 A. Avkhadiev and A. V. Sadofyev,	
{\it ``Chiral Vortical Effect for Bosons''},
  Phys. Rev. D96 (2017) 045015\
  arXiv:1702.07340 [hep-th].


\bibitem{sadofyev2}
``Chiral Vortical Effect For An Arbitrary Spin",
Xu-Guang Huang and A. V. Sadofyev, JHEP 1903 (2019) 08,
e-Print: arXiv:1805.08779 [hep-th] .

\bibitem{landsteiner}	
M.N. Chernodub, A. Cortijo, and K. Landsteiner,  
{\it ``Zilch vortical effect''},
  Phys. Rev. D98 (2018)  065016,
arXiv:1807.10705 [hep-th].


\bibitem{yamamoto}
Photonic chiral vortical effect,
N. Yamamoto,
Phys.Rev.D 96 (2017) 5, 051902,
1702.08886 [hep-th]

\bibitem{prokhorov1}
CVE for photons: black-hole vs. flat-space derivation
 G.Yu. Prokhorov,, O.V. Teryaev, V.I. Zakharov,
e-Print:
        2003.11119 [hep-th]

\bibitem{mitkin}
Zilch Vortical Effect, Berry Phase, and Kinetic Theory,
 Xu-Guang Huang, Pavel Mitkin, Andrey V. Sadofyev, Enrico Speranza(
e-Print:
        2006.03591 [hep-th].

\bibitem{vilenkin}
A. Vilenkin, 	{\it 
``Quantum Field Theory At Finite Temperature In A Rotating System''}
  Phys. Rev. D21 (1980) 2260.


  
\bibitem{sonsurowka}
 D. T. Son and  P. Surowka,	
{\it ``Hydrodynamics with Triangle Anomalies''}
    Phys. Rev. Lett. 103 (2009) 191601, 
arXiv:0906.5044 [hep-th]. 


\bibitem{reviews}
Kharzeev D., Landsteiner K., Schmitt A., Yee H.-Y,
Strongly Interacting Matter in Magnetic Fields //
Lect. Notes Phys. 2013 V. 873 P. 1-624

 

\bibitem{shevchenko}
A.V. Sadofyev, V.I. Shevchenko, and V.I. Zakharov,
{\it ``Notes on chiral hydrodynamics within effective theory approach''},
 Phys.Rev. D83 (2011) 105025,
arXiv:1012.1958 [hep-th]. 

\bibitem{stone}
 M. Stone and J. Kim, 	
{\it ``Mixed Anomalies: Chiral Vortical Effect and the Sommerfeld Expansion''},
  Phys. Rev. D98 (2018) 025012,
 arXiv:1804.08668 [cond-mat.mes-hall]. 

\bibitem{megias}
K. Landsteiner, Eu. Megias, and F. Pena-Benitez,
{\it ``Gravitational Anomaly and Transport''},
  Phys. Rev. Lett. 107 (2011) 021601,
 arXiv:1103.5006 [hep-ph].



\bibitem{megias1}
K. Landsteiner, E. Megias, and F. Pena-Benitez, 
{\it ``Anomalies and Transport Coefficients:The Chiral Gravito-Magnetic Effect''},
arXiv:1110.3615 [hep-ph].

\bibitem{wilzcek}
 S. P. Robinson and F. Wilczek,
{\it ``Relationship  between  Hawking  Radiation  and  Gravitational Anomalies''},
 Phys. Rev. Lett. 95 (2005) 011303;\\
 S. Iso, H. Umetsu, and F. Wilczek,
{\it ``Hawking Radiation from Charged Black Holes via Gauge and Gravitational
Anomalies''}, 
Phys. Rev. Lett.96, 151302 (2006);\\
S. Iso, H. Umetsu,  and F. Wilczek,
{\it Anomalies,  Hawking Radiations  and Regularity in Rotating Black Holes''}, 
Phys. Rev.D74, 044017 (2006).

 \bibitem{vainshtein}
A.I. Vainshtein, A.D. Dolgov, V. I. Zakharov, and I.B. Khriplovich,	
{\it ``Chiral Photon Current And Its Anomaly In A Gravitational Field''},
  Sov. Phys. JETP 67 (1988) 1326, Zh. Eksp. Teor. Fiz. 94 
(1988) 54-64.

\bibitem{lubanski}
J. Lubanski, Sur le spin des paticles elementaires, Physica, 8 (1) 44-52 (1941)\\
W. Pauli, Relativistic field theory of elementary particles, 
{\it Reviews of modern physics}, 13 (3)203 91941).

\bibitem{becattini2}
Polarization in relativistic fluids: a quantum field theoretical derivation
F. Becattini,  e-Print:
        2004.04050 [hep-th].

\bibitem{teryaev}
Prokhorov G., Teryaev O. V.
Anomalous current from the covariant Wigner function//
 Phys. Rev. 2018 V. D97 P. 076013-076019,
arXiv:1707.02491 [hep-th] .



\bibitem{becattini}
Becattini F.
Thermodynamic equilibrium with acceleration and the Unruh effect //
Phys. Rev.  (2018) V. D97 P.  085013-08519
arXiv:1712.08031 [gr-qc] 


 \bibitem{zubarev}
Zubarev D. N., Prozorkevich A. V.,
 Smolyanskii S. A., Derivation of nonlinear generalized
equations of quantum relativistic hydrodynamics //Theoret. and Math. Phys, 1979 V. 40 P. 
821-831. 


 \bibitem{prokhorov4}
Prokhorov G., Teryaev O., Zakharov V.I.	
Axial current in rotating and accelerating medium//
Phys. Rev. 2018 V. D98 P. 071901
arXiv:1805.12029 [hep-th].


\bibitem{prokhorov5}
Prokhorov G., Teryaev O., Zakharov, V.I.
``Unruh effect universality: emergent conical geometry from density operator'',
JHEP 2003 (2020) 137,  arXiv:1911.04545 [hep-th].


\bibitem{duff}
M. J. Duff, Supergravity 81, Proc. of 1st School on Supergravity, 
Ed. by S. Ferrara and J. G. Taylor, Cambridge Univ.  Press,  1982 
[see: Introduction to Supergravity, Moscow, Mir, 1985 (in Russian) 1. 




\bibitem{sasha}
Dolgov A.D., Zakharov V.I.,
``On conservation of the axial current in massless electrodynamics'',
Nuclear Physics B, 15 (1971) 525-540.

\bibitem{luttinger}
``Theory of Thermal Transport Coefficients''
J. M. Luttinger
Phys. Rev. 135, A1505.

 \bibitem{vilenkin2}	
``Parity Nonconservation and Rotating Black Holes",
A. Vilenkin, 
 Phys.Rev.Lett. 41 (1978) 1575-1577.\\
``Parity Violating Currents in Thermal Radiation",
A. Vilenkin,  Phys.Lett. 80B (1978) 150-152.

\bibitem{mertens}
A. Blommaert, Th. G. Mertens, H. Verschelde, and V. I. Zakharov,  
{\it ``Edge State Quantization: Vector Fields in Rindler''},
JHEP 1808 (2018) 196,
arXiv:1801.09910 [hep-th] .

\bibitem{vainshtein1}
 A.D. Dolgov, I.B. Khriplovich, A.I. Vainshtein, and V. I. Zakharov,	
{\it ``Photonic Chiral Current and Its Anomaly in a Gravitational Field''},
 Nucl. Phys. B315 (1989) 138.

\bibitem{prokhorov}
 G. Prokhorov, O. Teryaev, and V. Zakharov, 	
{\it ``Axial current in rotating and accelerating medium''},
  Phys. Rev. D98 (2018)  071901,
  arXiv:1805.12029 [hep-th].

\bibitem{yin}
M.A. Stephanov and Y. Yin, 
{\it "Chiral Kinetic Theory''},
Phys. Rev. Lett. 109 (2012) 162001б,
 arXiv:1207.0747 [hep-th] .




 \bibitem{hou}
 De-Fu Hou,  Hui Liu, Hai-can Ren,
{\it``A Possible Higher Order Correction to the Vortical Conductivity 
in a Gauge Field Plasma''},
    Phys. Rev. D 86 (2012) 121703,
arXiv 1210.0969 [hep-th].

\bibitem{Sorin:2016smp}
A.~Sorin and O.~Teryaev,
``Axial anomaly and energy dependence of hyperon polarization in Heavy-Ion Collisions,''
Phys. Rev. C \textbf{95} (2017) no.1, 011902
[arXiv:1606.08398 [nucl-th]].

\bibitem{Baznat:2013zx}
M.~Baznat, K.~Gudima, A.~Sorin and O.~Teryaev,
``Helicity separation in Heavy-Ion Collisions,''
Phys. Rev. C \textbf{88} (2013) no.6, 061901
[arXiv:1301.7003 [nucl-th]].

\bibitem{Ivanov:2020qqe}
Y.~B.~Ivanov,
``Global polarization in heavy-ion collisions based on axial vortical effect,''
[arXiv:2006.14328 [nucl-th]].

\bibitem{Baznat:2017jfj}
M.~Baznat, K.~Gudima, A.~Sorin and O.~Teryaev,
``Hyperon polarization in heavy-ion collisions and holographic gravitational anomaly,''
Phys. Rev. C \textbf{97} (2018) no.4, 041902
[arXiv:1701.00923 [nucl-th]].

\bibitem{Singha:2020qns}
S.~Singha [STAR],
``Measurement of global spin alignment of $K^{*0}$and $\phi$ vector mesons using the STAR detector at RHIC,''
[arXiv:2002.07427 [nucl-ex]].

\bibitem{Efremov:1981vs} 
A.~V.~Efremov and O.~V.~Teryaev,
``ON HIGH P(T) VECTOR MESONS SPIN ALIGNMENT,''
Sov. J. Nucl. Phys. \textbf{36} (1982), 557
JINR-P2-81-859.

\bibitem{Becattini:2020ngo}
F.~Becattini and M.~A.~Lisa,
``Polarization and Vorticity in the Quark Gluon Plasma,''
[arXiv:2003.03640 [nucl-ex]].

\bibitem{Luschevskaya:2018chr}
E.~V.~Luschevskaya, O.~V.~Teryaev, D.~Y.~Golubkov, O.~V.~Solovjeva and R.~A.~Ishkuvatov,
``Tensor polarizability of the vector mesons from $SU(3)$ lattice gauge theory,''
JHEP \textbf{11} (2018), 186
[arXiv:1811.02344 [hep-lat]].

\bibitem{Luschevskaya:2020aag}
E.~V.~Luschevskaya, O.~V.~Teryaev, R.~A.~Ishkuvatov and O.~E.~Solovjeva,
``Hadron Polarization in Strong Magnetic Field,''
Phys. Part. Nucl. Lett. \textbf{17} (2020) no.3, 289-295

\bibitem{Buividovich:2012kq}
P.~V.~Buividovich, M.~I.~Polikarpov and O.~V.~Teryaev,
``Lattice studies of magnetic phenomena in heavy-ion collisions,''
Lect. Notes Phys. \textbf{871} (2013), 377-385
[arXiv:1211.3014 [hep-ph]].

\bibitem{Buividovich2010}
P.~V.~Buividovich, M.~N.~Chernodub, D.~E.~Kharzeev, T.~Kalaydzhyan, E.~V.~Luschevskaya and M.~I.~Polikarpov,
``Magnetic-Field-Induced insulator-conductor transition in SU(2) quenched lattice gauge theory,''
Phys. Rev. Lett. \textbf{105} (2010), 132001
[arXiv:1003.2180 [hep-lat]].

\bibitem{Astrakhantsev:2019zkr}
N.~Y.~Astrakhantsev, V.~V.~Braguta, M.~D'Elia, A.~Y.~Kotov, A.~A.~Nikolaev and F.~Sanfilippo,
``Lattice Study of Electromagnetic Conductivity of Quark-Gluon Plasma in External Magnetic Field,''
[arXiv:1910.08516 [hep-lat]].

\bibitem{Artru:2008cp}
X.~Artru, M.~Elchikh, J.~M.~Richard, J.~Soffer and O.~V.~Teryaev,
Phys. Rept. \textbf{470} (2009), 1-92
[arXiv:0802.0164 [hep-ph]].

\bibitem{Gavrilova:2019jea}
M.~Gavrilova and O.~Teryaev,
``Rotation-invariant observables as Density Matrix invariants,''
Phys. Rev. D \textbf{99} (2019) no.7, 076013
[arXiv:1901.04018 [hep-ph]].




\bibitem{adler}
S. L. Adler,
{\it ''Recent Path Crossings with Roman and Anomalies'',}
    e-Print: 1910.04089 [hep-th].

\bibitem{sonzhitnitsky}
    D.T. Son
    and A. R. Zhitnitsky
{\it "Quantum anomalies in dense matter''}
        Phys. Rev.D 70 (2004) 074018, 
        hep-ph/0405216 [hep-ph].

\bibitem{zhitnitsky}
   M. A. Metlitski and  A. R. Zhitnitsky,
{\it "Anomalous axion interactions and topological currents in dense matter''},
        Phys. Rev. D 72 (2005) 045011,
        hep-ph/0505072 [hep-ph].

\bibitem{kirilin}
    V.P. Kirilin, 
    , A.V. Sadofyev, and V.I. Zakharov, 
{\it "Chiral Vortical Effect in Superfluid''},
        Phys. Rev. D 86 (2012) 025021,
        1203.6312 [hep-th].

\bibitem{teryaevzakharov}
O. V. Teryaev  and V  I. Zakharov, 
{\it "From the chiral vortical effect to polarization of baryons: A model''}        
        Phys. Rev. D 96 (2017) 9, 096023.


\end{thebibliography}
\end{document}